\documentclass[aps,prl,twocolumn,groupedaddress,showpacs,amsmath,amssymb]{revtex4}

\usepackage{graphicx}
\usepackage{dcolumn}
\usepackage{bm}

\begin{document}

\title{Oscillatory angular dependence of the magnetoresistance 
in a topological insulator Bi$_{1-x}$Sb$_{x}$}

\author{A. A. Taskin}
\author{Kouji Segawa}
\author{Yoichi Ando}
\affiliation{Institute of Scientific and Industrial Research, 
Osaka University, Ibaraki, Osaka 567-0047, Japan}


\begin{abstract}

The angular-dependent magnetoresistance and the Shubnikov-de Haas
oscillations are studied in a topological insulator
Bi$_{0.91}$Sb$_{0.09}$, where the two-dimensional (2D) surface states
coexist with a three-dimensional (3D) bulk Fermi surface (FS). Two
distinct types of oscillatory phenomena are discovered in the
angular-dependence: The one observed at lower fields is shown to
originate from the surface state, which resides on the
$(2\bar{1}\bar{1})$ plane, giving a new way to distinguish the 2D
surface state from the 3D FS. The other one, which becomes prominent at
higher fields, probably comes from the $(111)$ plane and is obviously of
unknown origin, pointing to new physics in transport properties of
topological insulators.

\end{abstract}



\maketitle

The three-dimensional (3D) topological insulator hosts a novel surface
state whose existence is guaranteed by a topological principle
characterized by the $Z_2$ invariant \cite{K1,K2,MB}, which makes it an
interesting playground for exploring the role of topology in condensed
matter \cite{SCZ1,K4,SCZ3,H1,H2,H3,Matsuda,AliY,Xue,QO,YiCui,Ong,SM,
Furu,TYN,Vish,DHL}. The key feature of the topological insulator is
that an odd number of massless, spin-helical Dirac cones comprise the 2D
surface state, whose gapless nature is protected by the time-reversal
symmetry. Such an intrinsically conducting surface is supported by an
insulating bulk, where the energy gap is created by a strong spin-orbit
coupling. To understand the macroscopic properties of this novel surface
state and to investigate the possibility of its device applications,
transport studies are obviously important. However, in actual samples of
topological insulators, there is always some bulk conductivity due to
residual carriers, and separating the contributions from 2D and 3D
states turns out to be challenging \cite{QO,YiCui,Ong}. 

Recently, we observed \cite{QO} strong de Haas-van Alphen (dHvA)
oscillations in high-quality bulk single crystals of Bi$_{1-x}$Sb$_x$
alloy in the ``insulating" regime (0.07 $\le x \le$ 0.22), which is the
first material to be known as a 3D topological insulator \cite{K2,H1}.
The dHvA oscillations signified a previously-unknown Fermi surface (FS)
with a clear 2D character that coexists with a 3D bulk FS \cite{QO}.
Since Bi$_{1-x}$Sb$_x$ is a 3D material, the observed 2D FS is naturally
assigned to the surface. In the present work, to specifically probe
low-dimensional properties of this material, we have extended our study
to the angular dependence of the magnetoresistance. This method was
successfully applied to the studies of quasi-2D organic conductors in
the 1980s, resulting in the discovery of the celebrated
angular-dependent magnetoresistance oscillations (AMRO)
\cite{AMRO1,AMRO2,AMRO3}, and later extended to quasi-one-dimensional
conductors \cite{AMRO4,AMRO5}. In this Letter, we present a detailed
study of the magnetotransport in a Bi$_{0.91}$Sb$_{0.09}$ single
crystal, where we have found oscillatory angular dependences in both the
resistivity $\rho_{xx}$ and the Hall resistivity $\rho_{yx}$. The
oscillatory magnetoresistance (MR) is obviously different from known
AMRO \cite{AMRO5}, and it is comprised of two components: one is a
peculiar manifestation of the 2D FS, and the other appears to be a
fundamentally new effect. The former provides a new way to distinguish
the contributions of the 2D FS, while the latter points to new physics
in topological insulators.

High-quality Bi$_{1-x}$Sb$_x$ crystals were grown from a stoichiometric
mixture of 99.9999\% purity Bi and Sb elements by a zone melting method
\cite{QO}. The resistivity was measured by a standard four-probe method
on a rectangular sample. In this Letter, we focus on a $x$ = 0.09 sample
in which the current was directed along the $C_1$ axis. Continuous
rotations of the sample in constant magnetic fields was used to measure
the angular dependence of the MR within two main crystallographic planes
($C_3$-$C_2$ and $C_3$-$C_1$). To observe the SdH oscillations,
$\rho_{xx}(B)$ and $\rho_{yx}(B)$ were also measured by sweeping $B$
between +14 and -14 T along a set of magnetic-field directions.

\begin{figure}\includegraphics*[width=7.5cm]{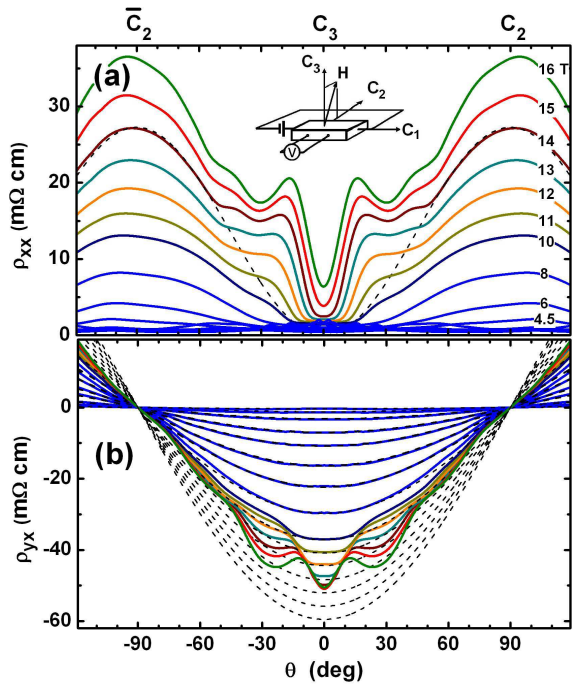}
\caption{(Color online) 
(a) Angular dependences of $\rho_{xx}$ measured in the trigonal-binary
($C_3$-$C_2$) plane in constant magnetic fields. 
The dashed line is the MR background $\sim (1-\cos^{2} \theta)$ for 
$B$ = 14 T.
Inset depicts the measurement geometry, where
$\theta$ = 0$^{\circ}$ corresponds to the $C_3$ axis.
(b) Angular dependences of $\rho_{yx}$ measured in the same conditions 
as in (a). The dashed lines are the expected angular dependences of the 
Hall effect plotted for all magnetic fields (0.1, 0.25, 0.5, 1, 2, 
3, 4.5, 6, 8, 10, 11, 12, 13, 14, 15, and 16 T).
}
\label{fig1}
\end{figure}

Let us start by presenting our main observations. Figure 1(a) shows the
angular dependences of the transverse MR measured in magnetic fields
rotated within the trigonal-binary ($C_3$-$C_2$) plane perpendicular to
the current. The magnetic-field strength, ranging from 0.01 T to 16 T,
were kept constant during each rotation. As clearly seen in Figs. 1--3,
pronounced oscillations of both $\rho_{xx}$ and $\rho_{yx}$ as a
function of the rotation angle $\theta$ appear for $B \gtrsim$ 1 T. Two
different types of oscillations can be distinguished: The first type
consists of oscillations appearing at lower fields (Figs. 1 and 2),
while the second one becomes prominent at higher fields (Figs. 1 and 3).
These data were taken immediately after the sample surface was refreshed
by a chemical etching \cite{note1}.

Figure 2 magnifies the ``low-field" oscillations of $\rho_{xx}(\theta)$,
which become prominent above 1 T. These oscillations are symmetric with
respect to the $C_{3}$ axis ($\theta$ = 0$^{\circ}$) and show a clear
tendency of shifting the peak positions closer to the center with
increasing magnetic field. It is important to notice that this strong
field dependence is very different from ordinary AMRO \cite{AMRO3} where
the angular positions of peaks are determined by commensurate motions of
electrons for certain field directions on warped FSs \cite{AMRO4, AMRO5}
and, thus, are field-independent. Peculiarly, the amplitude of observed
oscillations decrease with increasing field above $\sim$8 T and
disappear at higher fields. It turns out that the angular positions of
the outer-most peaks (with respect to $\theta$ = 0$^{\circ}$), which are
marked by arrows in Fig. 2, follow a simple geometrical law (inset of
Fig. 2) related to the response of the 2D electrons residing on the
$(2\bar{1}\bar{1})$ plane perpendicular to the $C_1$ axis, as will be
discussed later. Note that, because of the crystal symmetry, there are
three equivalent $C_1$ axes in the plane perpendicular to the trigonal
($C_3$) axis; hence, two of the three equivalent 2D FSs are at
$\pm$30$^{\circ}$ to the rotation plane. The $\rho_{xx}(\theta)$
measurements of the same sample in magnetic fields rotated within the
trigonal-bisectrix ($C_3$-$C_1$) plane (not shown) reveal essentially
similar angular dependences.

\begin{figure}\includegraphics*[width=7.5cm]{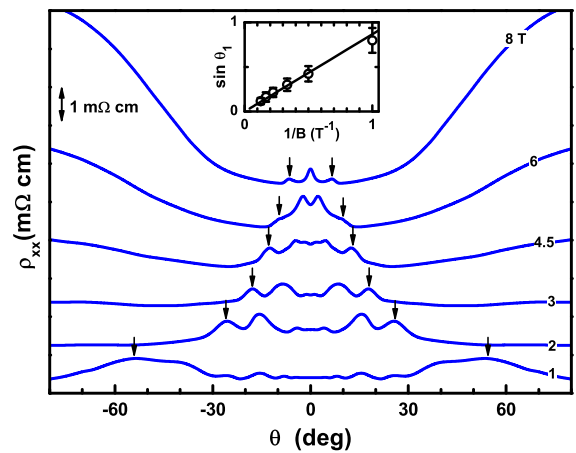}
\caption{(Color online) 
``Low-field" angular dependences of $\rho_{xx}$ shown in Fig 1(a).
Curves are shifted for clarity. Arrows mark the positions of the first
peaks, which appear as the field direction is rotated from the $C_{2}$
($\bar{C_{2}}$) to $C_{3}$ axis. Inset shows the dependence of
$\theta_1$ on $B$ plotted as $\sin \theta_1$ vs $B^{-1}$, which gives
evidence for the 2D-FS origin of the peaks.
}
\label{fig2}
\end{figure}

In contrast to these ``low-field' oscillations, the ``high-field"
angular dependence of the transverse MR is developing on a smooth
field-dependent background coming from the anisotropy of $\rho_{xx}(B)$
along the different axes. An example of its fitting for $B$ = 14 T is
shown by the dashed line in Fig. 1(a). Because of a large MR background
in strong magnetic fields, only largest peaks in $\rho_{xx}(\theta)$ can
be clearly seen in the raw data [Fig. 1(a)]. Subtracting the background
would help to reveal smaller peaks, but it is not necessary, because the
situation is much simpler in the Hall data as shown below.

Figure 1(b) shows the $\rho_{yx}(\theta)$ data, which also show
pronounced angular-dependent oscillations at high fields. The
``background " for $\rho_{yx}(\theta)$ is simply the angular dependence
of the Hall effect, $R_{H}B\cos\theta$, where $R_{H}$ is the Hall
coefficient. As can be clearly seen in Fig. 1(b), low-filed
$\rho_{yx}(\theta)$ data follow this expected angular dependence very
closely (we use $R_{H} = -37$ cm$^{2}$/C, obtained from the Hall
measurements), and the large deviation from this simple behavior is
observed only in magnetic fields above 10 T. Figure 3 shows ``pure"
oscillations in $\Delta \rho_{yx}(\theta)$ after subtracting the $R_{H}
B \cos \theta$ contribution from $\rho_{yx}(\theta)$. One can clearly
see a set of peaks, which are marked by short vertical ticks in Fig. 3.
They are symmetric with respect to the $C_{3}$ axis and show a rather
complicated magnetic-field dependence.

\begin{figure}\includegraphics*[width=7.5cm]{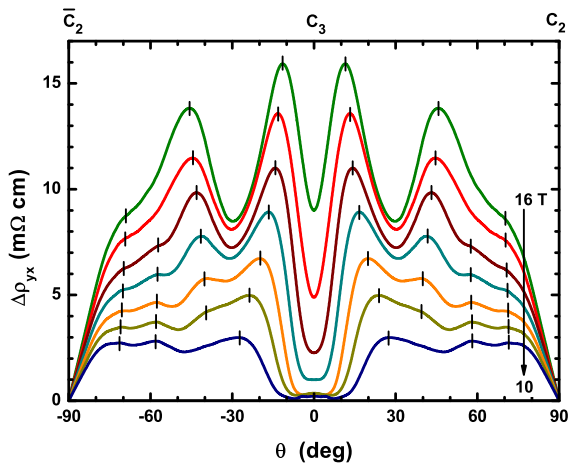}
\caption{(Color online)
``High-field" angular-dependent oscillations of $\Delta \rho_{yx}$
obtained by subtracting the expected $R_{H} B \cos \theta$ behavior from
the $\rho_{yx}$ data shown in Fig. 1(b). Ticks mark the positions of
distinguishable peaks.
}
\label{fig3}
\end{figure}

To understand the origin of the observed angular-dependent oscillations,
detailed knowledge of the FS is useful. We therefore measured the SdH
oscillations on the present sample for a series of magnetic-field
directions in two high-symmetry planes, as shown in Fig. 4(a) for the
$C_3$-$C_1$ plane. Those SdH oscillations consist of three fundamental
frequencies, which we call $f_1$, $f_2$, and $f_3$. How these
frequencies vary in the $C_3$-$C_1$ plane is shown in Fig. 4(b), which
is essentially consistent with what was observed with the dHvA
oscillations \cite{QO} and indicates the presence of both 2D and 3D FSs.
In the present sample, the radius of the circular 2D FS (given by $f_1$)
is $k_{F}$ = 4.15$\times$10$^{5}$ cm$^{-1}$ which is the same as in Ref.
\onlinecite{QO}, while the 3D FS is somewhat smaller, the semi-axes of
the three ellipsoidal 3D FSs (given by $f_2$ and $f_3$) are $a$ =
1.2$\times$10$^{6}$ cm$^{-1}$, $b$ = 8.5$\times$10$^{5}$ cm$^{-1}$, and
$c$ = 4.2$\times$10$^{5}$ cm$^{-1}$. Note that the 2D FS resides on the
plane perpendicular to the $C_1$ axis, while the 3D ellipsoids are
located at the $L$ points of the bulk Brillouin zone, with the longest
axes in the binary plane and tilted from the $C_1$ axis by
$\sim$5$^{\circ}$. The divergence of $f_1$ at $\theta=0^\circ$ seen in
Fig. 4(b) is of geometric origin and is a hallmark of the 2D FS, as can
be seen in the inset of Fig. 4(b).

The analysis of the SdH amplitude gives us the cyclotron mass $m_c$ and
the scattering time $\tau$. For the field direction very close to the
$C_3$ axis, the SdH oscillations coming from the 2D and 3D FSs can be
easily separated, and their fits to the standard Lifshitz-Kosevich
theory \cite{Schnbrg} at $\theta \approx$ 3$^{\circ}$ yield $m_c/m_e$ of
0.13 and 0.033 for the 2D and 3D FSs, respectively ($m_e$ is the free
electron mass). Note that the relatively large $m_c$ for the 2D FS
originates from its characteristic angular dependence $m_c = m_c^0/\sin
\theta$, where $m_c^0 = 0.0057 m_e$ is for the orbital motion in
perpendicular fields ($B \parallel C_1$), meaning that the 2D electrons
are actually extremely light. Corresponding Dingle plots (not shown)
give Dingle temperatures $T_D$ of about 5.5 K and 7 K for the 2D and 3D
FSs, which imply the mean free paths of $l^{2D} \approx$ 150 nm and
$l^{3D} \approx$ 16 nm, respectively \cite{note2}. This difference is
not surprising, if one remembers the Dirac nature \cite{ShGu1,ShGu2} of
the 2D surface states and that they are topologically protected against
spin-conserving backscattering.

\begin{figure}\includegraphics*[width=8.7cm]{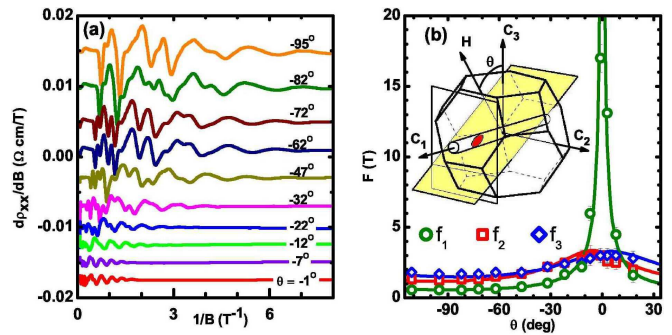}
\caption{(Color online) 
(a) SdH oscillations measured in the $C_3$-$C_1$ plane. For better
visibility, $d\rho_{xx}/dB$ is calculated after a smooth background MR
is subtracted, and the curves are shifted. (b) $\theta$-dependences of
the three fundamental frequencies obtained from the Fourier transform of
the SdH oscillations. Inset shows a schematic picture of the bulk
Brillouin zone, in which the 2D FS is crossed by a plane perpendicular
to the magnetic-field direction.
}
\label{fig4}
\end{figure}

We now discuss the origin of the ``low-field" part of the
angular-dependent oscillations in Bi$_{0.91}$Sb$_{0.09}$. A key insight
comes from the recognition that the 3D FSs in our sample enter the
quantum limit (where all the electrons condense into the first Landau
level) with magnetic fields of only 1--3 T for any direction, while the
necessary magnetic field to bring the 2D FS into the quantum limit
diverges as the field direction is rotated toward the $C_3$ axis. This
means that, in sufficiently high magnetic field, the 3D electrons will
always remain in the quantum limit, while the 2D electrons can always be
brought from the quantum limit back to the regime where higher Landau
levels are populated. This will lead to the oscillating behavior of
$\rho_{xx}$ due to the 2D FS alone when the magnetic field is rotated
toward the $C_3$ axis. Remembering that the resistivity oscillates as
$\rho_{xx} \sim \cos[2\pi (F/B_{\rm eff} + \gamma)]$ with $B_{\rm eff} =
B \sin \theta$, a maximum in $\rho_{xx}$ occurs when $F/B_{\rm eff} +
\gamma$ = $n$ with integer $n$. Hence, one can understand that the Fermi
level is crossed by the $n$-th Landau level of the 2D FS at the specific
angle $\theta_n$ given by 
\begin{equation} 
\theta_{n} = \arcsin \left ( \frac{F/B}{n-\gamma} \right ) ,
\end{equation} 
where $F$ is the frequency and $\gamma$ is the phase of the SdH
oscillations. According to this model, the outer-most peaks marked by
arrows in Fig 2 are coming from the crossing of the Fermi level by 
the first Landau level ($n$ = 1) at a given magnetic-field strength, 
so their angles should correspond to $\theta_1$. 
Actually, the inset of Fig. 2 demonstrates that Eq. (1) fits the data
reasonably well. If $F$ is assumed to be 0.65 T based on the SdH result
\cite{note3}, the fit in Fig 2 yields $\gamma$ of 0.25, suggesting a
non-zero Berry phase associated with the 2D FS \cite{Mikitik,Kim}.

\begin{figure}\includegraphics*[width=6cm]{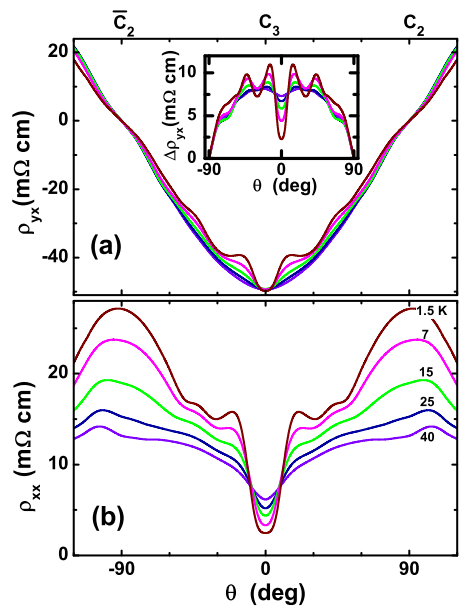}
\caption{(Color online) 
(a) Angular-dependent oscillations of $\rho_{yx}$ within the $C_3$-$C_2$ 
plane in the constant field $B$ = 14 T measured at 1.5, 7, 15, 25, and 
40 K. Inset shows oscillations of $\Delta \rho_{yx}$ after subtraction of
the $R_{H} B \cos \theta$ term.
(b) Angular-dependent oscillations of $\rho_{xx}$ within the $C_3$-$C_2$ 
plane measured in the same conditions as in (a).
}
\label{fig5}
\end{figure}

Finally, let us discuss the origin of the ``high-field" peaks shown in
Figs. 1 and 3. Here, an important feature is that the amplitude of these
peaks weakens as the magnetic field is rotated away from the $C_3$ axis,
which is somewhat reminiscent of the behavior of the ordinary AMRO in
quasi-2D systems if the conduction planes lie perpendicular to the $C_3$
axis \cite{AMRO1,AMRO2,AMRO3}. Note that this is different from what is
observed for the ``low-field" oscillations, which are originating from
the surface state residing on the $(2\bar{1}\bar{1})$ plane ($\perp
C_{1}$). Thus, it is probable that the ``high-field" oscillations are
coming from the (111) plane ($\perp C_{3}$), where surface states are
seen in photoemission \cite{H1,H2,H3,Matsuda} and tunneling \cite{AliY}
experiments. Another distinguishable feature of the ``high-field"
oscillations is that they survive up to rather high temperatures as can
be seen in Fig. 5, where temperature dependences of both
$\rho_{yx}(\theta)$ and $\rho_{xx}(\theta)$ are shown. Even at 40 K
there are still visible traces of oscillations (see the inset of Fig.
5), while the ``low-field'' SdH and angular-dependent oscillations
disappear around 20 K. In spite of some similarities to the quasi-2D
AMRO, the peak positions of the ``high-field" oscillations apparently
shift with the magnetic field, which is not expected for the ordinary
AMRO. Moreover, the existence of a finite coupling between conduction
planes is essential for the quasi-2D AMRO \cite{AMRO5}, but there is no
such inter-plane coupling for the surface states as long as the crystal
is thick enough. Therefore, the observed ``high-field" angular
oscillations are a new phenomenon apparently specific to topological
insulators, and it is possible that they are associated with a coupling
between the surface and the bulk states.

In conclusion, we found two different types of oscillatory phenomena in
the angular-dependent MR of a topological insulator
Bi$_{0.91}$Sb$_{0.09}$. One type is observed at lower fields and
provides a new way to distinguish the 2D FS. The other one, which
becomes prominent at higher fields, probably comes from the $(111)$
plane and points to new physics in transport properties of topological
insulators.

\begin{acknowledgments}
We thank V. M. Yakovenko for discussions. This work was supported by
JSPS (KAKENHI 19674002 and 20030004) and AFOSR (AOARD 08-4099 and
10-4103). 
\end{acknowledgments}


\begin{thebibliography}{99}
\bibitem{note1}
We observed weaker oscillations in the same sample before etching,
corroborating their surface origin.
\bibitem{note2}
The scattering time $\tau$ (= $\hbar / 2 \pi k_{B}T_D$) and 
the Fermi velocity $v_F$ (= $\hbar k_{F} / m_{c}$) gives the 
mean free path $\ell$ = $v_{F} \tau$.
In the present sample, $v_{F}^{3D} \approx$ 7.6$\times$10$^{6}$ cm/s is
an order of magnitude smaller than $v_{F}^{2D} \approx$
8.4$\times$10$^{7}$ cm/s.
\bibitem{note3}
The SdH frequency of the 2D FS is 0.57 T when $B$ is perpendicular to
the 2D plane, but it becomes 0.65 T when $B$ is 30$^{\circ}$ tilted, as
is the case for two of the three equivalent 2D FSs in $B \parallel C_2$.
\end{thebibliography}
\end{document}